\documentclass[twocolumn,english]{article}
\usepackage[T1]{fontenc}
\usepackage[latin9]{inputenc}
\usepackage{graphicx}
\usepackage{underscore}
\usepackage{caption}
\usepackage{subcaption}
\usepackage[top=2cm, bottom=2cm, left=2cm, right=2cm]{geometry}
\usepackage{setspace}

\usepackage{savesym}
\usepackage{amsmath}
\savesymbol{iint}
\usepackage{txfonts}
\restoresymbol{TXF}{iint}

\makeatletter
 \usepackage{overcite}
\date{}

\makeatother

\usepackage{babel}
\begin{document}
%
%
%

\twocolumn[
\title{A general, efficient and robust method for calculating free energy difference between
systems}
\author{Asaf Farhi$^{\dagger}$\\
\small{Physics Department Weizmann Institute of Science, Rehovot 76100, Israel}\\}
\medskip
\maketitle
\textbf{Calculating free energy differences is a topic of
substantial interest and has many applications including chemical reactions, molecular
docking and hydration, solvation and binding free energies which are used in computational drug discovery.
However, in the existing practices in relative free energy calculations the implementation is rather complicated, the simulated hybrid system can suffer from small phase space overlap and there remain the challenges of robustness and automation. Here an
efficient and robust method, that enables a wide range of comparisons, will be introduced, demonstrated and compared.
In this method instead of artificially transforming one system into the other to perform the calculation, each system is transformed into its replica with the different long range energy terms relaxed, which is \emph{inherently correlated} with the original one, in order to eliminate the partition function difference arising from these terms.
Then, since each transformed system can be treated as non interacting systems, the remaining difference in the (originally highly complex) partition function will cancel out.
}
\medskip]

\let\thefootnote\relax\footnote{$^{\dagger}$ asaf.farhi@gmail.com}

Calculating free energy differences between two physical systems,
is a topic of substantial current interest. A variety of advanced
methods and algorithms have been introduced to answer the
challenge, both in the context of Molecular Dynamics and Monte
Carlo simulations 
\cite{chipot2007free,zuckerman2011equilibrium,frenkel1996understanding,shirts2007alchemical,binder2010monte}.
 Applications of these methods include calculations of
binding free energies \cite{kollman1993free,deng2009computations,woods2011water},
free energies of hydration \cite{straatsma1988free}, free energies
of solvation \cite{khavrutskii2010computing}, transfer of a
molecule from gas to solvent \cite{shirts2007alchemical}, chemical reactions \cite{jorgensen1987priori} and more.
Binding free energy calculations are of high importance since they
can be used for molecular docking \cite{meng2011molecular} and
 in drug discovery\cite{jorgensen2009efficient}. Free energy methods are extensively
used by various disciplines and the interest in this field is
growing - over 3,500 papers using the most popular free energy
computation approaches were published in the last decade, with the
publication rate increasing $\sim17\%$ per year
\cite{chodera2011alchemical}.

Free energy difference between two systems can be calculated using
equilibrium methods (alchemical free energy calculations) and non
equilibrium methods. In equilibrium methods a hybrid system is
used to transform system $A$ into $B$, usually with the
transformation $H_{hybrid}=\lambda
H_{A}+\left(1-\lambda\right)H_{B},\,\lambda=[0,1]$ . In these methods, first the
intermediates ($\lambda$s) that interpolate between the systems
are selected and the hybrid system is simulated at these
intermediates and average values are calculated. Then, using these
values, the free energy difference is calculated. The commonly
used methods include Bennett Acceptance Ratio
\cite{bennett1976efficient}, Weighted Histogram Analysis Method
\cite{kumar1992weighted}, Exponential Averaging/ Free Energy
Perturbation \cite{zwanzig1954high} and Thermodynamic Integration (ThI)
\cite{frenkel1996understanding,kirkwood1935statistical,straatsma1991multiconfiguration}.
In non equilibrium methods the work needed in the process of
switching between the two Hamiltonians is measured. These methods
include Jarzynski relation \cite{jarzynski1997nonequilibrium}
(fast growth is one of its applications \cite{hendrix2001fast})
and its subsequent generalization by Crooks \cite{crooks1999path}.
Most of the applications mentioned above can be tackled from a
different direction using methods which measure the free energy as
a function of a reaction coordinate. These methods include
Adaptive Biasing Force \cite{darve2001calculating} and Potential
Mean Force \cite{kirkwood1935statistical} (fast growth also
belongs to this type of methods).

Calculating binding free energies is fundamental and has many applications. In particular it has potential to advance
the field of drug discovery which has to cope with new challenges.
In the last years the number of innovative new molecular entities
(for pharmaceutical purposes) has remained stable at $5-6$ per
year. This situation is especially grim when taking into account
the continual emergence of drug-resistant strains of viruses and
bacteria. Virtual screening methods, in which the $10^{60}$ possible molecules are filtered out, play a large role in modern drug discovery efforts.
However, there remains the challenge of selecting the candidate molecules out of the
still very large pool of molecules in reasonable times.
Equilibrium methods show great potential in enabling the
computation of binding free energies with reasonable computational
resources. In these methods instead of simulating the
binding processes directly, which would require a simulation many
times the lifetime of the complex, the ligand is transmuted into another through intermediate, possibly nonphysical
stages \cite{chodera2011alchemical}. Then if the free energy difference between the ligands in the two environments are calculated \cite{Detailed}, the binding free energy difference between the two ligands can be calculated (this cyclic calculation is called the Thermodynamic Cycle). 
\begin{figure}[h]
 \centering
\includegraphics[width=7cm]{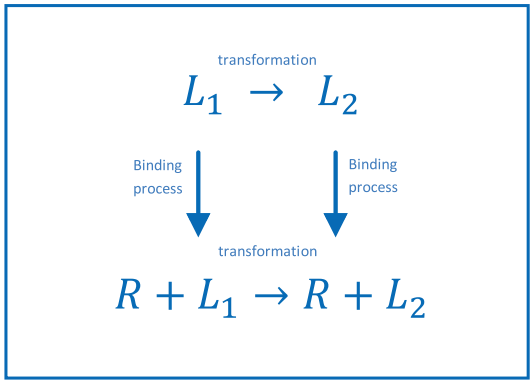}
 \caption{Scheme of the free energy differences in the calculation of binding free energy in the existing methods}
 \label{fig:binding}
\end{figure}
In Fig \ref{fig:binding} a scheme of the free energies in the calculation of binding free energies in the existing methods is presented ($L_1,L_2$ and $R$ represent the ligands and receptor respectively).

Free energy calculation methods already have successes in
discovering potent drugs\cite{jorgensen2009efficient}. However, despite the continuing progress in the field from the original concepts,
 the methods have restrictions which prevent them from being automatic and limit their use in
computational drug design \cite{chodera2011alchemical}.
A naive calculation of the free energy difference using ThI can be performed as follows: 
 \begin{align}
\triangle F_{A\rightarrow B}\left(\beta_{1}\right)=\int_0^1\frac{dF_{A\rightarrow B}\left(H_{hybrid}\left(\lambda\right)\right)}{d\lambda}d\lambda=\\
\int_0^1\int\frac{\left[H_{B}(\Omega)-H_{A}(\Omega)\right]e^{-\beta_{1}\left[\lambda H_{B}(\Omega)+\left(1-\lambda\right)H_{A}(\Omega)\right]}d\Omega}{Z(\lambda)}d\lambda
\end{align}
It can be seen that at $\lambda=1$ for example $H_{A}$ does not affect the systems' behavior but
its energy values are averaged over, which can result in large magnitudes of the integrated function. Thus, when the systems have low phase space overlap
there are significant changes in the integrated function and large variance and hence large computational cost. This is especially dominant
when the systems have different covalent bond description which results in a very low phase space overlap.
Many approaches and techniques have been introduced to address the challenges in the field. These include the topologies for simulating the hybrid system (single and dual topology), soft core potentials, the notion of decoupling atoms and more. Common sampling techniques to overcome energy barriers include Temperature and Hamiltonian Replica Exchange methods.
However, the implementation of equilibrium methods is notoriously difficult to implement correctly \cite{shirts2012best}.
Such complications arise for example from the fact that in the hybrid system there is one set of coordinates that describes both systems simultaneously and hence the hybrid system usually has to be designed. In addition each type of topology has small phase space overlap in one aspect \cite{pearlman1994comparison}. Also, since both systems interact simultaneously with the environment the behavior of the intermediate systems cannot be predicted.
Moreover,  since the process of transforming one system into the other is non physical and different for each
comparison, the choice of intermediates for the hybrid system remains a challenge.

Temperature Integration (TeI) was suggested in \cite{FarhiTemperature} as an
efficient method to calculate free energy differences. While TeI
has many advantages it cannot be directly applied to molecular simulations
in which bond stretching is allowed and to Molecular Dynamics.
This is since the use of the capping in the covalent bond terms
(denoted as $E_{cap}$ in TeI ) will result in a phase transition which will
result in impractical simulations in the canonical ensemble
 and since MD simulations at very high temperatures (suggested in
TeI) are impractical due to the very high velocities. In addition, the phase space overlap between the compared systems is rather low since, effectively, all the energy terms are relaxed.  
 Here, based on TeI, a general method that solves these three challenges will be presented. 
This method also addresses the current challanges in the field that were stated before.

The method was derived from principles of statistical physics, independently of the methodology in the field mentioned above. In section the method will be introduced in a logical developmental order and will be linked in places of similarity to the relevant literature. In section the derivation will be put in context of the methodology in the field and will be divided into its ingredients. This section is also aimed for practitioners that are interested in the decription of the method in the terminology of the field (the article is thus aimed both for physicists and practitioners). As will become clear, the independent ingredients of the method are: a different approach to topology with a relevant sampling technique, unified approach to soft core potentials and a minor improvement in the decoupling scheme. It is emphasized that the main ingredient of the method is that instead of simulating a hybrid system, two separate systems are simulated (related to topology). 

\section*{The method}
The idea is to transform the systems $A$ and $B$ (between which
the free energy difference is calculated) into two systems that
have the same partition function up to factors that can be calculated. In the transformation
the energy terms are lowered (or unchanged) and as a result the corresponding
partition functions value will change monotonously, which will enable
a simple selection of intermediates for the calculation of free
energy difference.
 In this process in case there are covalent bonds terms (represented by harmonic potential),
 their strength will stay constant in order to avoid phase transition (molecular modeling includes covalent bond, bond angle, dihedral angle, electric and VDW potentials \cite{phillips2005scalable,hess2008gromacs}).
For the systems $A$,$B$ the following Hamiltonians as a function
of $\lambda$ are defined:
\begin{equation}
H_{A/B}\left(\lambda\right)=\lambda
H_{_{{A_{nc}/B_{nc}}}}+H_{{}{A_c/B_c}}
\end{equation}
where $H_{_{{A_{nc}/B_{nc}}}}$ denote the non covalent potential
terms and $H{{}_{A_{c}/B_{c}}}$ denote the covalent bond terms,
with the corresponding partition functions:
\begin{equation}
\tilde{Z}_{A/B}\left(\lambda\right)=\int
e^{-\beta\left(\lambda
H_{_{{A_{nc}/B{nc}}}}+H_{{A_c/B_c}}\right)}d\Omega
\end{equation}
It has to be noted that the integration over momentum is not
presented since it yields known terms and can be factored out.

Since even at $\lambda\rightarrow0$ the energy terms
 that diverge at $r=0$ are still dominant and cause a difference
 between the partition functions of the two systems,
capping is used in the non covalent terms (if $E>E_{cap}$ $E=E_{cap}$).
Thus, the partition functions at $\lambda\rightarrow0$ are equal
up to factors that originate from the covalent bonds that are
different.
The proposed calculation of the free energy difference between the
two systems is legitimate only if the choice of the capping energy
has a negligible effect on the partition function value of each of
the two systems at $\lambda=1$. The Hamiltonian with the capping
 in all the non covalent energy terms denoted by
$H'$ is written as follows:
\begin{equation}
H'_{A/B}\left(\beta,\lambda\right)=\lambda
H'_{_{{A_{nc}/B_{nc}}}}+H_{{}{A_c/B_c}}
\end{equation}
The requirement stated above can be written explicitly as follows:
\begin{equation}\label{cutoffA}
ln\tilde{Z}_{A/B}\left(\beta,\lambda=1,H'\right)\backsimeq
ln\tilde{Z}_{A/B}\left(\beta,\lambda=1,H\right)
\end{equation}

It can be seen that at $\lambda=0$:
\begin{equation}
ln\tilde{Z}_{A}\left(\beta,\lambda=0,H'\right)-ln\tilde{Z}_{B}\left(\beta,\lambda=0,H'\right)=\\
\end{equation}
\begin{equation}
\Delta ln\tilde{Z}_{{A_c\rightarrow B_c}}\left(\beta,0,H'\right)
\end{equation}
where $ln\tilde{Z}_{A_{c}/B_{c}}\left(\beta,\lambda=0,H'\right)$
are defined as follows:
\begin{equation}
\tilde{Z}_{A_{c}/B_{c}}\left(\beta,\lambda=0,H'\right)=\int
e^{-\beta H_{A_{c}/B_{c}}}d\Omega
\end{equation}
In order for the capping to have a negligible effect on the partition
functions at $\lambda=1$ it has to be set to a value that satisfies:
\begin{equation}
e^{-\frac{E_{cap}}{kT}}\ll 1
\end{equation}
Thus at $\lambda$ values satisfying $e^{-\frac{\lambda E_{cap}}{kT}}\ll 1$  the corresponding interactions, including the steric,
become transparent \cite{Detailed}. It is noted that the capping has a negligible effect on the partition function of realistic molecular
systems independently of the atom pairs. This is since each atom has an average
potential energy of typical value of $k_{B}T$ above the average energy (the capping energy can be set relative to the the average pair total long range energy) and since the denisty of states for $E>E_{cap}$ is very low (dual effect). It has been shown in the context of MC that there is a tradeoff when choosing $E_{cap}$ (behavior of the function vs. accuracy) \cite{FarhiThesis} and that values of $~5$kCal/mol \cite{FarhiThesis} enable accurate free energy calculations. 
 The fact that the force is zero in some of the range is legitimate in MD simulations since the particles have thermal velocities and they are affected by other potentials (also there exist other potentials without force at certain distances e.g the VdW and the Coulomb potentials at large distances). It is noted that a switching function between the standard long range potential and the flat potential is needed in order have continuity in the derivative that will enable the integration over the equations of motion in MD to be valid. This has been developed independently and implemented in MD with a cubic switching function and an energetically inaccessible capping which validates its use in the context of MD for high energetic values \cite{buelens2012linear}. Thus a unified approach is suggested that uses quadratic switching function and capping energy that is accessible and has a negligible effect on the free energy as a soft core technique.

From here on $H'$ will be used in all the $\tilde{Z}$ terms:
\begin{equation}
\tilde{Z}_{A/B}\left(\beta,\lambda\right)\triangleq\tilde{Z}_{A/B}\left(\beta,\lambda,H'\right)
\end{equation}
When the systems have the same degrees of freedom the free energy
difference can be written as follows:
\begin{equation}
F_{A\rightarrow B} (\beta) = -k_BT \left[\ln{Z_B(\beta)} -
\ln{Z_A(\beta)} \right] \backsimeq \nonumber\\
\end{equation}
\begin{equation}
-k_BT\big{[} \ln{\tilde{Z}_B(\beta,1)} -
\ln{\tilde{Z}_B(\beta,0)} + \ln{\tilde{Z}_A(\beta,0)} \nonumber\\
\end{equation}
\begin{equation}
 \Delta
\ln{\tilde{Z}_{{
B_c\rightarrow A_c}}(\beta,0)}-\ln{\tilde{Z}_A(\beta,1)} \big{]}  = \nonumber\\
\end{equation}
\begin{equation}
\int_0^1 \langle H'_{B_{nc}}\rangle d\lambda - \int_0^1 \langle
H'_{A_{nc}}\rangle d\lambda - k_BT\Delta \ln{\tilde{Z}_{{
B_c\rightarrow A_c}}(\beta,0)}
\end{equation}
using equations (\ref{cutoffA}), the earlier
definition of $\Delta ln\tilde{Z}_{{B_c\rightarrow
A_c}}\left(\beta,0\right)$ and the following identity:
\begin{equation}
\ln\tilde{Z}_{A/B}\left(\beta,1\right)-ln\tilde{Z}_{A/B}\left(\beta,0\right)=\int_0^1\frac{dln\tilde{Z}_{A/B}\left(\beta,\lambda\right)}{d\lambda}d\lambda\\
\end{equation}
Assuming the factor $\Delta ln\tilde{Z}_{B_{c}\rightarrow A_{c}}\left(\beta,0\right)$, that can be different than zero just
in molecular simulations in which bond stretching is allowed, is
known, the calculation of the free energy difference can thus be
performed. 
Thus, the systems are simulated at the
\emph{same} temperature and the molecule stays in its covalently bounded state (which enables MD simulations that are not practical at very high temperatures due to the short time steps needed 
and modelling that includes bond strecthing respectively).

When the systems, between which the free energy difference is
calculated, have rugged energy landscape, one can use techniques
such as H-REMD/H-PT (Hamiltonian Replica Exchange MD/ Hamiltonian Parallel Tempering, variant of Parallel Tempering/Replica Exchange \cite{ferrenberg1988new,hansmann1997parallel,earl2005parallel}) to alleviate sampling problems \cite{fukunishi2002hamiltonian}. In this technique the system is simulated at a set of $\lambda$s and
exchanges of configurations between them are performed every
certain number of steps. Thus, the systems at the low $\lambda$s,
that can cross energetic barriers, help the system of interest to
be sampled well. This technique, even though is highly efficient,
introduces another dimension of sampling since simulations of
replicas of the system at a set of $\lambda$s are performed at
each intermediate of the hybrid system (sampling the dimensions of
$\lambda$ that interpolates between the systems and of $\lambda$
of the replicas that are used for the equilibration). Here, the
simulations at the different $\lambda$s will be used \emph{also}
to calculate the free energy difference by integration and
the need for another sampling dimension is eliminated.

It has to be mentioned that the less energy terms are multiplied by $\lambda$,
the more correlated the systems at $\lambda=0$
and at $\lambda=1$ are, resulting in a much shorter calculation.
In particular, keeping the
 dihedral and bonding angles terms constant can result in less intermediates and
 shorten the simulation time even more. Thus the energy terms can be separated into short range (covalent bond, bond angle, dihedral angle and improper dihedral angle) and long range (electric and VDW). In the existing practices they are called bonded interactions and non bonded interactions and it was decided to use these names in order to emphasize the fact that the potentials link up to 3rd nearest neighbor and all atoms respectively which have implication later on. 
The Hamiltonians can be written as follows:
 \begin{equation}
H'_{A/B}\left(\lambda\right)=\lambda
H'_{_{{A_{lr}/B_{lr}}}}+H_{{}{A_{sr}/B_{sr}}}
\end{equation}
where $H_{_{{A_{lr}/B_{lr}}}}$, $H_{{}{A_{sr}/B_{sr}}}$ denote the long and short range energy terms respectively.
Thus the four most dominant types of interactions are kept constant in the transformation. In the next section it will be explained how to compensate for the corroesponding factor $\Delta ln\tilde{Z}_{{B_{sr}\rightarrow
A_{sr}}}\left(\beta,0\right)$ .

In order to equilibrate the entire system the energy terms that are not multiplied by $\lambda$ can be written as follows:
 \begin{equation}
 H\rightarrow f\left(\lambda\right) H_,\,\,\,\,\, f\left(\lambda\right)=\left\{ \begin{array}{cc}\lambda & \lambda\geq\lambda_{eq}\\\lambda_{eq} & \lambda<\lambda_{eq}\end{array}\right.
 \end{equation}
 where $\lambda_{eq}$  denote the minimal $\lambda$ for equilibration in the H-REMD procedure.
Here we transformed only up to $\lambda_{eq}$ in order to have minimal transformation.
Thus the H-REMD procedure is in its original form in the range $\lambda=[1,\lambda_{eq}]$ and the systems at $\lambda=[\lambda_{eq},0]$ can be simulated separately.
It is noted that the H-REMD procedure, which is in the range $\lambda=[1,\lambda_{eq}]$, does not depend on the compared molecule and the comparison is associated with the simulations in the range  $\lambda=[\lambda_{eq},0]$.
Covalent bond and bond angle energy terms may not need equilibration (multiplication by $\lambda$) as they are not expected to be associated with rugged energy landscape. 

The partition function value at $\lambda=0$
 can in many cases be calculated, enabling the calculation of
absolute free energy value (up to a factor that is determined by
the choice of length unit and the number of DOF that cancels out in comparisons).

\subsection*{The remaining free energy difference}
We now turn to show how to treat the remaining
free energy difference. This will be done in two contexts: in the context of direct
free energy calculation between the systems and in the context of
the Thermodynamic Cycle.
The context of direct calculation can be used to calculate the free
energy difference between two different molecules in the same environment. This direct calculation can be applied to chemical reactions in which the molecules are different and to the best of our knowledge has not been suggested. This may be used to calculate free energies of chemical reactions which have been possible only with a combination of Quantum Mechanical calculations. 
 The context
of the Thermodynamic Cycle is for calculating the free energy difference
between the same system (keeping its covalent structure) in two states.
This has many applications which include solvation and binding free
energies and is here presented in a systematic derivation that enables
to keep constant (almost) all the short range energy terms and even some long range terms, enabling to maintain high
phase space overlap between the compared systems.

In order to perform the direct calculation we can switch to relative coordinates (coordinates of atoms relative to other colvalently bounded atoms) and then to spherical coordinates. 
Thus, the direct calculation of the partition function of the molecule only
with the short range energy terms can be performed by integration.

 The molecule is first divided into elements of standard covalent bonds,  bond junctions 
and of more complex structures that include molecular rings. Since
each of the short range potentials depends on one independent variable, the 
the integration in each element is independent of the others. Thus the
integrals can be performed separately and then multiplied to yield
the partition function and hence the free energy difference.
%

For a system of two atoms with a covalent bond term represented by
a harmonic potential (as is in molecular simulation software) the
partition function can be readily calculated \cite{SM}.
In addition, the partition
function of complex structures at
$\lambda=0$ can often be calculated numerically. Other internal bonding energy terms
can also be included in these numerical integrations (and also not be multiplied by $\lambda$).
Alternatively, systems with similar complex structures can be compared as in other methods, eliminating the need for these calculations.

The factor $\Delta ln\tilde{Z}_{B_{c}\rightarrow
A_{c}}\left(\beta\right)$ that originates from the different
covalent description can be written as follows:
\begin{equation}
\Delta ln\tilde{Z}_{B_{c}\rightarrow
A_{c}}\left(\beta\right)=\sum_{i}ln\tilde{Z}_{B_{c_{i}}}-\sum_{j}ln\tilde{Z}_{A_{c_{j}}}
\end{equation}
where $i$ and $j$ denote the elements that are different in the systems $A$ and $B$
respectively. Similar treatment can be applied to the bond and dihedral angle potentials \cite{Detailed}. Dihedral angle energy terms that will introduce complexity can be relaxed in the transformation.

When using the method in the context of the Thermodynamic Cycle we can also switch to relative coordinates and then to spherical coordinates. We can identify a separation point between the part that is common and different between the compared molecules. If at the separation point the potentials will depend seperately on the independent variables we can decouple the common and different partition functions. We can fulfill this requirement if at the seperation point there will be one dihedral potential and no improper dihedral potential terms (other can be relaxed in the transformation). 
Thus it can be written:
\begin{equation}
Z\rightarrow Z_{\mathrm{common\, int}}Z_{\mathrm{diff\, non\, int}}
\end{equation}
where $Z_{\mathrm{common\, int}}$ represents the partition function
of the common part between the compared molecules, $Z_{\mathrm{diff\, non\, int}}$ denotes the partition function
of the different part of the molecule and the arrow symbolizes the transformation in which
the different long range energy terms are relaxed.
 The free energy that is associated with the the different part will cancel out independently of its complexity in the Thermodynamic Cycle since it does not interact with the environement and therefore the same.
\begin{figure}[h]
 \centering
\includegraphics[width=8cm, height=6cm]{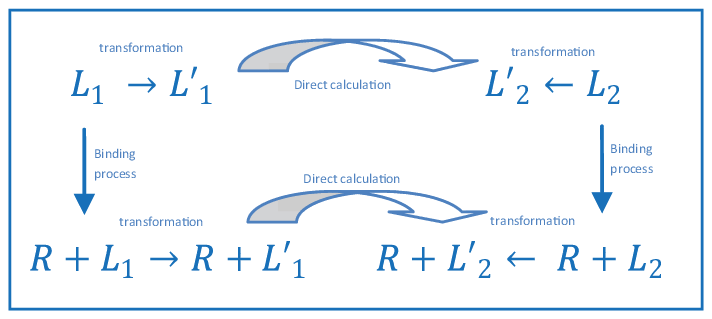}
 \caption{Scheme of the free energy differences in the novel method}
 \label{fig:bindingh}
\end{figure}
In Fig \ref{fig:bindingh} a scheme of the free energies in the novel method is presented. The ligand $L_1$/ $L_2$ is transformed into its replica with the long range interactions (VDW and electric) of the different atoms between the systems relaxed  $L_1'$/ $L_2'$. The free energy difference between  $L_1'$ and  $L_2'$ does not need to be calculated since it cancels out ($\Delta F_{L'_1\rightarrow L'_2}=\Delta F_{R+L_1'\rightarrow R+ L'_2}$). The free energy difference arising from the short range interactions is in many cases the dominant part of the total free energy difference.
 Hence, for realistic molecular systems, only a capping energy has to be set on the long range potential terms and each system should be transformed into its replica with the long range energy terms of the atoms that are different between the molecules relaxed \cite{Detailed}. 
This can be applied also for the comparison of molecules with different molecular ring content. Thus, simple comparison between any two systems with the same degrees of freedom can be performed.
We write the Hamiltonian in its final form:
 \begin{equation}
H'_{A/B}\left(\lambda\right)=\lambda
H'_{_{{A_{dlr}/B_{dlr}}}}+f(\lambda) H_{{}{A/B}}
\end{equation}
where $H_{_{{A_{dlr}/B_{dlr}}}}$ denote the different long range energy terms.
The corresponding calculation of solvation/binding free energies in the case of systems that do not necessitate H-REMD procedure can thus be written:
  \begin{equation}
\Delta F_{A_{\mathrm{solvation/binding}}\rightarrow B_{\mathrm{solvation/binding}}}=\int_0^1\left\langle H'_{B_{dlr}}\right\rangle -\int_0^1\left\langle H'_{A_{dlr}}\right\rangle 
\end{equation}
\begin{equation}
+\int_0^1\left\langle H'_{A_{\mathrm{solvated/bounded}_{dlr}}}\right\rangle -\int_0^1\left\langle H'_{B_{\mathrm{solvated/bounded}_{dlr}}}\right\rangle 
\end{equation}
 
We emphasize the fact that in this standard context there is no need to calculate anything besides the calculation mentioned in the equation above.
The method has been demonstrated and compared in the context of the direct calculation on systems of a molecule of two atoms in a spherical potential \cite{SM}. 

\section*{Discussion}
A novel method for calculating free energy difference
between systems is presented. This method can be used to calculate
the free energy difference between a wide range of systems that
have the same degrees of freedom and is highly efficient, robust and
simple. In this method there are no large penalties in the running time
that originate from the dissimilarity between the systems. Moreover, since the
$\lambda$s used in the H-REMD/H-PT procedure are also used as
intermediates in the calculation of free energy difference, a
convergence for systems with rugged energy landscape is achieved
without introducing another sampling dimension. Both in the
calculation of the integral for the free energy difference and in
the H-REMD/H-PT procedures, the chosen intervals between the
$\lambda$s have to be smaller where the internal energy varies
significantly, in the free energy difference calculation in order
to have good sampling of the function and in H-REMD in order to
maintain optimal acceptance rates. Thus, no additional unnecessary
$\lambda$s have to be sampled. 

In equilibrium methods the hybrid system has one set of coordinates that 
specifies the configurations of the two systems so the hybrid system often needs to be designed. 
In addition when the hybrid Hamiltonian involves potentials with more complicated lambda dependence, their derivatives may have to be calculated. The method has the advantage of 
simplicity since the simulations are performed only on the two (almost) original systems (separate simulations) and the need to relate between the compared systems is eliminated. 

Using this method (preceded by virtual
screening filtering) an automated free energy calculation
that will result in the best candidates may be
 performed. It is noted that the method can be applied to calculate free energies of systems composed of weakly interacting subsystems (e.g spin system in a magnetic field with weak coupling).
There is a provisional patent pending that includes the content of this paper. Prof. D. Harries
and Prof. G. Falkovich are acknowledged for the useful comments.

\section*{Supplementary Material}
\subsection*{Demonstration and comparison}
In order to demonstrate the method it was used with all its
ingredients in MC simulation to calculate the free energy
difference between the systems $A$ and $B$ and this value was
compared to the one calculated by numerical integration. Then, in
order to asses the efficiency of the method, the free energy
difference between the systems was calculated using ThI
 combined with H-PT (in MC simulation) and the running
time of the two methods was compared.

The compared systems are composed of a molecule of two atoms in which one atom is fixed to
the origin and the second one is bound to the first by a covalent
bond. The second atom in each system is in a $\theta$ dependent
potential (in spherical coordinates), containing $\theta^{-12}$
term to represent the VDW repulsive term used in molecular
modeling. The potential barrier was chosen to be of typical value of systems with tens of atoms, having rugged energy landscape. The covalent bond length difference was chosen to represent systems with few different atom lengths- see the next section for more details (the values of the pairs of spring constant and covalent bond length were taken from molecular simulation software). 
The following potential and parameters
were used:
\begin{equation}
V_{A}=\frac{1}{2}k\left(r_{12}-r_{eq_{A}}\right)^{2}-5.5k_{b}T\cdot\left[sin\left(4\theta\right)-\frac{10^{-8}}{\left(\theta-\frac{3\pi}{8}\right)^{12}}\right]
\end{equation}
\begin{equation}
V_{B}=\frac{1}{2}k\left(r_{12}-r_{eq_{B}}\right)^{2}+5k_{b}T\cdot\left[sin\left(4\theta\right)+\frac{10^{-8}}{\left(\theta-\frac{5\pi}{8}\right)^{12}}\right]
\end{equation}
$$r_{eq_{A}}=2.1\,\AA,r_{eq_{B}}=1.3\,\AA, k_{A}=123\,kCal/Mol\AA^{2}$$
$$k_{B}=428\,kCal/Mol\AA^{2}, E_{cut\, off}=7\,kCal/Mol$$

Here we used the partition function of two atoms with a covalent bond term represented by
a harmonic potential that can be written as follows:

\begin{equation}
Z\left(\beta\right)=\frac{1}{l^{3}}\int_{-\infty}^{\infty}e^{-\beta
k\left(r-d\right)^{2}}d^{3}x=\frac{4\pi}{l^{3}}\int_{0}^{\infty}e^{-\beta
k\left(r-d\right)^{2}}r^{2}dr=
\end{equation}
\begin{equation}
\frac{\pi\left[\left(2d^{2}\beta k+1\right)\left(erf\left(d\sqrt{\beta k}\right)+1\right)+2\sqrt{\pi}e^{-\beta kd^{2}}d\sqrt{\beta k}\right]}{4l^{3}\left(\beta k\right)^{3/2}}
 \end{equation}
where $l$ is an arbitrary length.
This was used in eq. $\left(17\right)$ to calculate the free energy difference between the systems with the long range interaction relaxed. 

The comparison of the method to the numerical integration yielded
exactly the same results. The running time of the calculation of
the free energy difference performed by the two methods was
compared and yielded a factor of $\sim30$ in favor of the novel
method. This factor originates from the extra sampling dimension and
the larger number of intermediates needed in ThI
 combined with H-PT.
 \begin{figure}[h!]
 \centering
\includegraphics[width=8cm]{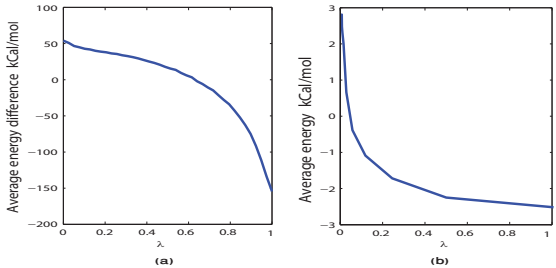}
 \caption{Integrated functions as a function of $\lambda$ (a) Thermodynamic Integration (b) The novel method}
 \label{fig:Combined}
\end{figure}
In Fig \ref{fig:Combined} the functions integrated in the two methods are plotted.
It can be seen that the difference in magnitude of the integrated function in the novel method is much smaller than
 this in ThI (factor of $\sim40$).
 \begin{figure}[h!]
 \centering
\includegraphics[width=8cm]{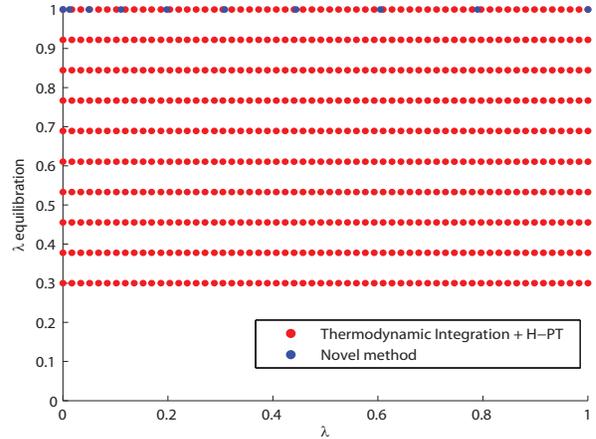}
 \caption{Systems simulated in both methods}
 \label{fig:Grid}
\end{figure}
In Fig \ref{fig:Grid} a scheme of the systems simulated in the two methods is presented (each point represents a simulation).

The dissimilarity between the systems that grows with the number of different particles, increases \emph{both} the number of intermediates (due to a much larger difference in magnitude) and the number of simulation steps (increased variance) as compared with these in the novel method. The difference in the covalent bond description, that reduces the correlation between the
systems significantly (the penalties are also not bounded by the capping energy) and has the most dominant effect, has a completely negligible computational cost in the novel method.
Thus, the efficiency is increased in 3 multiplicative dimensions.
It is here to remind that while the method is highly efficient, its biggest advantage is that it enables many comparisons 
of molecules with different connectivity. This is since two random molecules are not likely to have the same connectivity.

\subsection*{Correspondence between the toy model and realistic systems in the comparison to Thermodynamic Integration}

	In the case of realistic compared systems in which the molecules differ in the covalent bond length of one atom (usually when comparing two systems with different connectivity there are many such differences), when the long range energy terms of the atoms that are different between the molecules are relaxed (disregarding the equilibration procedure both in the existing methods and in the novel method for simplicity), the comparison of the methods will yield results that are very similar to the toy model. $\\*$
 This is since $\left\langle H_B-H_A \right\rangle (\lambda)$ in Thermodynamic Integration will be mainly affected by the changes between the systems. Thus, the functions integrated over in the toy model give good estimation to the ones in the comparison of the realistic systems mentioned above.
Since the value of the functions integrated by Thermodynamic Integration is dominated by the covalent bond change (the rest of the difference in the energy terms is negligible as compared with it) it can be written:
\begin{equation}
\left\langle H_B-H_A\right\rangle(\lambda)|_{realistic}\cong \left\langle H_B-H_A\right\rangle(\lambda)|_{toy\,model}\\* 
\end{equation}
Now we turn to show the correspondence in the novel method. We denote the energy terms of the atoms that are different between the realistic compared systems by $H_{A_d/B_d}$. Since these terms are the only ones in the integrated function it can be written: $\\*$
\begin{equation}
\left\langle H_{A/B}\right\rangle(\lambda)|_{realistic}=\left\langle H_{A_d/B_d}\right\rangle(\lambda)|_{realistic}\thicksim\left\langle H_{A/B}\right\rangle (\lambda)|_{toy\,model}\\*
\end{equation}
In this case there will be similar values since the energy values of the non covalent energy terms are bounded by $E_{cap}$ and thus the average value is of typical value of $E_{capping}$. Here the covalent bond energy term is notf included in $H_{A_d/B_d}$. The parameters in the toy model for the covalent bond lengths and strengths are realistic and were taken from simulation software.
\onecolumn 
\section*{Appendix}
{
\section*{Explanation of the capping}
\begin{figure}[H]
 \centering
\includegraphics[width=15cm]{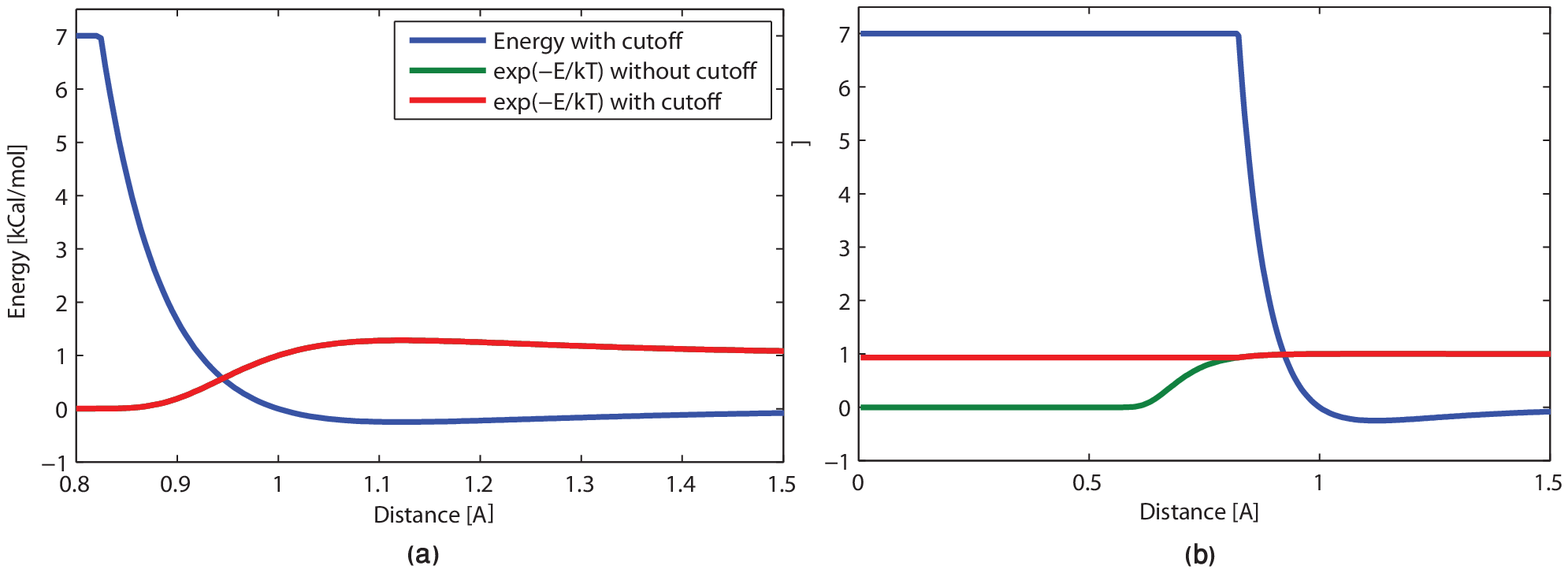}
 \caption{Energy and $\exp(-E/kT)$ as a function of distance for the potential $r^{12}-r^{-6}$ with $E_{cap}$=7kCal/mol (a) at $\lambda=1$ (b)  at $\lambda=0.01$}
 \label{fig:cutoff}
\end{figure}
In Fig \ref{fig:cutoff} energy and $\exp(-E/kT)$ as a function of distance for the potential $r^{12}-r^{-6}$ are plotted at $\lambda=1$ and at $\lambda=0.01$. It can be seen that the cap energy has a negligible effect on the probability distribution at $\lambda=1$ and that at small $\lambda$s the capping of the energy enables the equation of the partition functions.
It is noted that the capping has a negligible effect on realistic molecular
systems independently of the atoms whose long range potential is being
relaxed. This is since each atom has an average potential energy of
typical value of $k_{B}T$ and since the denisty of states for $E>E_{\mathrm{cutoff}}$
is very low (dual effect). In addition we mention that the force in
this range is zero which is legitimate for MD simulations since each
atom has a thermal velocity and is also affected by other potentials (this case also exist for VdW and coulomb potentials at large distances). It is noted that a switching function between the standard long range potential and the flat potential is needed in order to ensure continuity in the derivative that will enable the integration over the equations of motion in MD to be valid.
In addition, the capping of the enegy removes the limitation in the existing
methods in which the electric potential terms have to be removed before
the Lennard Jones terms in order to avoid divergence.
\section*{The remaining free energy difference}

We now turn to show how to treat the calculation of the remaining
free energy difference. This will be done in the context of direct
free energy calculation between the systems and in the context of
the Thermodynamic Cycle.

The context of direct calculation can be used to calculate the free
energy difference between two different molecules. This direct calculation
is applicable to chemical reactions in which the molecules are different
and to the best of our knowledge has not been suggested. The context
of the Thermodynamic Cycle is for calculating the free energy difference
between the same system (keeping its covalent structure) in two states.
This has many applications which include solvation and binding free
energies and is here presented in a systematic derivation that enables
to keep constant (almost) all the short range energy terms and even some long range potentials, enabling high
phase space overlap between the compared systems. While this was derived independently, it has similarities with the derivation in \cite{boresch2003absolute} which froms the basis of the relevant state of the art techniques. As compared to the existing practices, here more terms can remain constant in the transformation (e.g long range potentials between the different atoms).

\subsection*{Direct calculation}

The calculation of the free energy factors is based on the following
change of variables:

\[
d\Omega=\prod_{i=1}^{n}d\mathbf{r'_{i}}=d\mathbf{r'_{1}}\prod_{i=2}^{k}d\mathbf{r{}_{i}}\prod_{i=k+1}^{n}d\mathbf{r{}_{i}}=\prod_{i=1}^{k}d\mathbf{r'_{i}}\prod_{i=k+1}^{n}r_{i}^{2}sin\theta_{i}drd\theta d\phi
\]

where $\mathbf{r_{i}}=\mathbf{r'_{i}}-\mathbf{r'_{f\left(i\right)}}$,$r'_{f\left(i\right)}$
is the location of another atom that is bounded to atom $i$ (left
for choice), $1..k$ denote the similar atoms and $k+1,...,n$ denote
the different atoms between the compared molecules.

Next it will be assumed that the part in the molecule that is different
includes only short range interactions and in the beginning of each
sub-system there is one dihedral terms and the improper dihedral term
is relaxed.
The molecule is first divided into elements of standard covalent bonds,  bond junctions 
and of more complex structures that include molecular rings. Since
each of the short range potentials depend on one independent variable the 
the integration in each element is independent of the others. Thus the
integrals can be performed separately and then multiplied to yield
the partition function. We write these free energy factors explicitly:
\subsection*{Covalent bond}

The partition function of a covalent bond in Spherical Coordinates can be written as follows:
\begin{equation}
Z_c=\int e^{-\frac{\beta k_c\left(r-d\right)^{2}}{2}}r^{2}dr=\frac{\pi^2\left[\left(2d^{2}\beta k_c+1\right)\left(erf\left(d\sqrt{\beta k_c}\right)+1\right)+2\sqrt{\pi}e^{-\beta k_cd^{2}}d\sqrt{\beta k_c}\right]}{l^{3}\left(\beta k_c\right)^{3/2}}
 \end{equation}
where $l$ is an arbitrary length ($l^3$ cancels out in comparisons). 

\subsection*{Two Bonds Junctions}
When considering the case of a Linear/Bent molecular shapes, it can
be seen that in most cases when varying the bond angle, the rest of
the molecule moves as a rigid body. Since the spherical coordinates
representation includes three independent variables, varying a bonding
angle is decoupled from all the other degrees of freedom of the molecule.
Hence the calculation of free energy factors, which arise from bonding
potential terms that are different between the compared molecules,
is straightforward.
One of the most common bonding angles potentials is the following:
\begin{equation}
V_{b}\left(\theta\right)=\frac{1}{2}k_{\theta}\left(\theta-\theta_{0}\right)^{2}\end{equation}
So the integration over the corresponding degree of freedom can be written as:
\begin{equation}
Z_{b}=\int e^{-\beta V_{b}}d\Omega=\int e^{-\frac{\beta k_{\theta}}{2}\left(\theta-\theta_{0}\right)^{2}}sin\theta d\theta=\end{equation}
\begin{equation}
\frac{1}{2 \sqrt{\beta k_{\theta}}}e^{-i \theta_{0}-\frac{1}{2\beta k_{\theta}}} \sqrt{\frac{\pi }{2}} \left(i \text{Erf}\left[\frac{i-\theta_{0}\beta k_{\theta}+\beta k_{\theta} \pi }{\sqrt{2\beta k_{\theta}}}\right]+\text{Erfi}\left[\frac{1+i \theta_{0}\beta k_{\theta}}{ \sqrt{2\beta k_{\theta}}}\right]-i e^{2 i \theta_{0}} \left(\text{Erf}\left[\frac{i+\theta_{0}  k_{\theta}}{\sqrt{2\beta k_{\theta}}}\right]-i \text{Erfi}\left[\frac{1+i\beta k_{\theta} (\pi-\theta_{0} )}{\sqrt{2\beta k_{\theta}}}\right]\right)\right)
\end{equation}
This expression is real for positive and real values of $k_{\theta},\beta$ and $\theta_{0}$.

When varying a dihedral angle, the potential term value depends on
the orientation of first bond (which determines the axis from which
the dihedral angle is measured). However, since the integration has
to be performed over all the range $\left[0,2\pi\right]$, varying
the $\phi$ angle will yield a factor which is independent of the
location of the first bond. Thus, the integration does not depend on
the direction of the first bond and is straightforward.
The commonly used dihedral angles potential is of the following type:
\begin{equation}
V_{d}\left(\phi_{ijkl}\right)=k_{\phi}\left(1+cos\left(n\phi-\phi_{s}\right)\right)\end{equation}
The integration over this degree of freedom yields the following result:
\begin{equation}
Z_{d}=\int e^{-\beta V_{d}}d\Omega=\int e^{-\beta k_{\phi}\left(1+cos\left(n\phi-\phi_{s}\right)\right)}d\phi=2\pi e^{-\beta k_{\phi}}I_{0}\left(\beta k_{\phi}\right)\label{eq:dihedral}\end{equation}
where $I_{0}\left(\beta k_{\phi}\right)$ is Bessel function of the
first kind at $\beta k_{\phi}$ which is defined as follows:
\begin{equation}
I_{0}\left(x\right)=\sum_{l=0}^{\infty}\frac{\left(-1\right)^{l}}{2^{2l}\left(l!\right)^{2}}x^{2l}\end{equation}

\subsection*{Three or more Bonds Junctions}
Molecule shapes can include monomer that splits into more than one
monomer. Such examples are the trigonal planar, tetrahedral trigonal
pyramidal etc.
In these cases there is coupling between the monomers which will necessitate
numerical integration. This can be performed using the Spherical law
of cosines which can be written as follows:
\begin{equation}
cos\left(\theta_{12}\right)=cos\left(\theta_{1}\right)cos\left(\theta_{2}\right)+sin\left(\theta_{1}\right)sin\left(\theta_{2}\right)cos\left(\Delta\phi\right)\label{eq:Spherical law of cosines}\end{equation}
where $\theta_{1},\theta_{2}$ denote the bond angles of two bonds
with respect to the principal bond and $\theta_{12},\Delta\phi$ denote
the bond angle and the difference in $\phi$ angle between these two
bonds respectively.
Usually in these cases there is one dihedral angle energy term which
is between one of the bonds, the principal bond and a previous bond.
Since the integration over the other degrees of freedom yields a factor
that is independent of the value of $\phi$, the integrations can
be treated as decoupled.
Thus, the integration for the case of one monomer that splits into
two can be written as follows:
\begin{equation}
Z=\int e^{-\beta\left(V_{b}+V_{d}\right)}d\Omega=Z_{d}Z_{b}\end{equation}
where \[
Z_{d}=2\pi e^{-\beta k_{\phi}}I_{0}\left(\beta k_{\phi}\right)\]
and
\begin{equation}
Z_{b}=\int e^{-\frac{\beta}{2}\left[k_{1}^{\theta}\left(\theta_{1}-\theta_{1}^{0}\right)^{2}+k_{2}^{\theta}\left(\theta_{2}-\theta_{2}^{0}\right)^{2}+k_{12}^{\theta}\left(\theta_{12}-\theta_{12}^{0}\right)^{2}\right]}sin(\theta_{1})sin(\theta_{2})d\theta_{1}d\theta_{2}d\phi_{2}
\end{equation}
For the general case it can be written as follows:

\begin{equation}
Z_{b}=\int\prod_{i}e^{-\frac{\beta}{2}k_{i}^{\theta}\left(\theta_{i}-\theta_{i}^{0}\right)^{2}}\prod_{i>j}e^{-\frac{\beta}{2}k_{ij}^{\theta}\left(\theta_{ij}-\theta_{ij}^{0}\right)^{2}}\prod_{i}sin\theta_{i}d\theta_{i}\prod_{i\geq2}d\phi_{i}
\end{equation}
where $\theta_{ij}$ can be calculated from \eqref{eq:Spherical law of cosines}.
In case there are any energy terms that introduce complexity they can be relaxed in the transformation.

These free energy factors can be substituted in:
\begin{equation}
\Delta F_{B_{dsr}\rightarrow A_{dsr}}=k_{B}T\left[\sum_{i}lnZ_{B_{c_{i}}}+\sum_{i}lnZ_{B_{b_{i}}}+\sum_{i}lnZ_{B_{d_{i}}}-\left(\sum_{i}lnZ_{A_{c_{i}}}+\sum_{i}lnZ_{A_{b_{i}}}+\sum_{i}lnZ_{A_{d_{i}}}\right)\right]
\end{equation}
to give the remaining free energy difference. 

Thus, we can write in terms of the partition functions:

\[
Z\rightarrow Z_{\mathrm{common\, int}}\prod_{i=1}^{l}Z_{c_{i}}\prod_{i=1}^{m}Z_{d_{i}}\prod_{i=1}^{p}Z_{2b_{i}}\prod_{i=1}^{q}Z_{3b_{i}}\prod_{i=1}^{r}Z_{\mathrm{complex}_{i}}
\]

where $Z_{\mathrm{common\, int}}$ represents the partition function
of the common part between the compared molecules that is interacting
with the environment and $Z_{c_{i}}$ and $Z_{d_{i}}$ represent the
$i$th covalent bond and dihedral angle partition function respectively.
$Z_{2b_{i}}$ and $Z_{3b_{i}}$ represent the $i$th two bond and
three or more bond junctions respectively and $Z_{\mathrm{complex}_{i}}$
represents the $i$th complex structure partition function. The arrow
represents the transformation $\lambda=1\rightarrow0$.
\subsection*{The context of the Thermodynamic Cycle}

For the standard case in which the Thermodynamic Cycle is used the
change of variables takes the form:

\[
d\Omega=\prod_{i}^{n}d\mathbf{r'_{i}}=d\mathbf{r'_{1}}\prod_{i=2}^{k}d\mathbf{r{}_{i}}d\mathbf{r{}_{k+1}}\prod_{i=k+2}^{n}d\mathbf{r{}_{i}^{'}}=
\]

\[
d\mathbf{r'_{1}}\prod_{i=2}^{k}d\mathbf{r{}_{i}}r_{k+1}^{2}sin\theta_{k+1}dr_{k+1}d\theta_{k+1}d\phi_{k+1}\prod_{i=k+2}^{n}d\mathbf{r{}_{i}}
\]

This change of variables is motivated by the fact that the part that
is different between the molecules includes only short range interactions
and in the beginning of the different part there is one dihedral angle
and the improper dihedral term is relaxed.

Thus, we can write in terms of the partition functions:

\smallskip{}

\framebox{\parbox[c][0.8\height]{10cm}{%
\[
Z\left(\mathbf{r'_{1}},\mathbf{r{}_{2}},..,\mathbf{r_{n}}\right)\rightarrow Z_{\mathrm{common\, int}}\left(\mathbf{r'_{1}},\mathbf{r{}_{2}},..,\mathbf{r_{k}}\right)Z_{\mathrm{diff\, non\, int}}\left(\mathbf{r_{k+1,...,}r_{n}}\right)
\]
}}

\smallskip{}

where $Z_{\mathrm{diff\, non\, int}}$ denotes the partition function
of the different part of the molecule that does not interact with
the environment and the arrow symbolizes the transformation in which
the long range energy terms are relaxed.

In the case of totally different molecules it can thus be written:

\[
Z\rightarrow Z_{\mathrm{diff\, non\, int}}
\]

Here we made use of the fact that if the potential in the separation
point between the common and different parts of the molecule depends
on the spherical variables then there is decoupling between the common
and different parts of the molecule. It is noted that $Z_{\mathrm{diff\, non\, int}}$
cancels out in the Thermodynamic cycle.

In case there are improper dihedral term or more than one dihedral
term at the separation point (non common), we can simply relax them
in the transformation. Thus $Z_{\mathrm{diff\, non\, int}}$ \emph{does
not} need to be calculated and it can include any short range potentials
including dihedral and improper dihedrals. This treatment for the
usual case in which the Thermodynamic Cycle is used demonstrates that
the method is general and can be automated. For clarity we summarize:
free energy calculations in equilibrium methods are usually performed
in the context of the Thermodynamic cycle. Since the partition function
can simply be splitted to the common and different parts and the free
energy associated with $Z_{\mathrm{diff\, non\, int}}$ cancels out
in the Thermodynamic cycle, there is no need to calculate the free
energy difference. It is noted that the long range terms between terms the different atoms can remain constant. In addition it might be that if the the long range terms between the different atoms and the common environement (e.g water in binding)  are kept constant it may have a small effect on the free energy.  
\section*{The Thermodynamic cycle of the novel method}

\begin{figure}[H]
 \centering
\includegraphics[width=10cm]{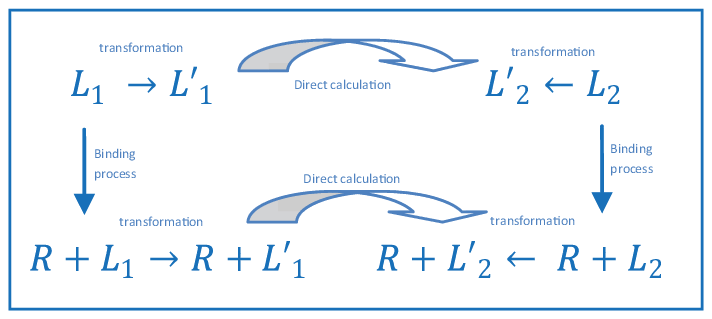}
 \caption{Scheme of the free energy differences in the novel method}
 \label{fig:bindingh}
\end{figure}
In Fig \ref{fig:bindingh} a scheme of the free energies in the novel method is presented. The ligand $L_1$/ $L_2$ is transformed into its replica with the long range interactions (VDW and electric) of the different atoms between the systems relaxed  $L_1'$/ $L_2'$. The free energy difference between  $L_1'$ and  $L_2'$ can be directly calculated but does not need to be since it cancels out ($\Delta F_{L'_1\rightarrow L'_2}=\Delta F_{R+L_1'\rightarrow R+ L'_2}$). The free energy difference arising from the direct calculation is in many cases the dominant part of the total free energy difference.
It can be written as follows:

$$\Delta F_{L_1\rightarrow R+L_1}-\Delta F_{L_2\rightarrow R+L_2}=\Delta F_{L_1\rightarrow L'_1}+\Delta F_{L'_1\rightarrow L'_2}-\Delta F_{L_2\rightarrow L'_2}-\left(\Delta F_{R+L_1\rightarrow R+ L'_1}+F_{R+L_1'\rightarrow R+ L'_2}-\Delta F_{R+L_2\rightarrow R+L'_2}\right)$$
\begin{equation}
=\Delta F_{L_1\rightarrow L'_1}-\Delta F_{L_2\rightarrow L'_2}-\left(\Delta F_{R+L_1\rightarrow R+ L'_1}-\Delta F_{R+L_2\rightarrow R+L'_2}\right)
\end{equation}
Thus, the relative binding free energy can be calculated.
This scheme is also relevant for relative free energy of solvation in which $R$ will represent the solvent instead of the receptor.

\section*{Use of the method for comparing two molecular systems}

As an example the simple molecules Benzoic acid and Toluene, that can be compared with this method, are presented. 

\begin{figure}[h!]
 \centering
\includegraphics[width=4cm]{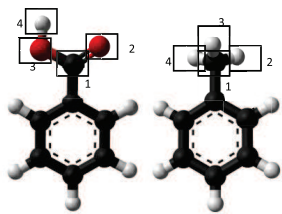}
 \caption{Benzoic acid (left) and Toluene (right) molecules}
 \label{fig:example}
\end{figure}

Calculating the free energy difference can be easily done in the method by relaxing only the long range (VDW and electric) energy terms of the atoms that are different between the molecules (atoms 2-4) in the transformation A/B to A'/B' by multiplying (only) them by $\lambda=1 \rightarrow 0$, decoupling their long range interactions from the rest of the molecule (and all their interactions with the environment). Now the partition function of each system can be separated into the complex PF of ring+atom $1$ which is identical between the systems (cancels out) and other simple decoupled PFs (direct calculation) or PF of the different part (TC) . The remaining difference in the free energy can be easily calculated by integration in the direct calculation and in the thermodynamic cycle it does not have to be calculated (cancels out). This
can be applied also for the comparison of molecules with different 
molecular ring content (e.g Cyclobutane and Benzene or larger molecules differing by such entities).  Thus, simple comparison between any two systems with the same degrees of freedom can be performed.

\twocolumn 

}

\begin{spacing}{0.5}
\bibliographystyle{unsrt}
{\footnotesize
\bibliography{bib12}}
\end{spacing}

\end{document}